\documentclass{optica-article}

\journal{opticajournal} 

\articletype{Research Article}

\usepackage{lineno}

\begin{document}

\title{Quantum enhanced mechanical rotation sensing using wavefront photonic gears}

\author{Ofir Yesharim,\authormark{1,*,$\dagger$} Guy Tshuva,\authormark{2,$\dagger$} and Ady Arie\authormark{1}}

\address{\authormark{1}School of Electrical Engineering, Iby and Aladar Fleischman Faculty of Engineering,
Tel Aviv University, Ramat Aviv 69978, Tel Aviv, Israel\\
\authormark{2}Raymond and Beverly Sackler School of Physics and Astronomy, Tel Aviv University, Ramat Aviv 69978, Tel Aviv, Israel\\}

\email{\authormark{*}{ofiryesharim@gmail.com}} 
\noindent\authormark{$\dagger$} equal contribution


\begin{abstract*}
Quantum metrology leverages quantum correlations for enhanced parameter estimation. Recently, structured light enabled increased resolution and sensitivity in quantum metrology systems. However, lossy and complex setups impacting photon flux, hinder true quantum advantage while using high dimensional structured light. We introduce a straightforward mechanical rotation quantum sensing mechanism, employing high-dimensional structured light and a compact high-flux (45,000 coincidence counts per second) N00N state source with N=2.  The system utilizes two opposite spiral phase plates with topological charge of up to $\ell$=16  that convert mechanical rotation into wavefront phase shifts, and exhibit a 16-fold enhanced super-resolution and 25-fold enhanced sensitivity between different topological charges, while retaining the acquisition times and with negligible change in coincidence count. Furthermore, the high photon flux enables to detect mechanical angular acceleration in real-time. Our approach paves the way for highly sensitive quantum measurements, applicable to various interferometric schemes.

\end{abstract*}

\section{Introduction}

Quantum metrology is a promising avenue in the second quantum revolution, where outperforming classical metrology systems is a key goal \cite{10.1116/5.0007577}. This enhanced performance is characterized by an improvement in resolution, as well as improved sensitivity, that can approach the Heisenberg uncertainty limit, which scales as $1/N$ instead of $1/\sqrt{N}$ for classical sensors, where $N$ is the number of photons. Within the field of quantum metrology, two main resources are pursued: Continuous variables squeezed states and discrete photon number N00N states \cite{10.1116/5.0007577}. While the potential of quantum metrology using these resources is known \cite{PhysRevLett.110.181101}, several challenges need to be addressed for wide usage of these systems, including: Low flux of entangled photons, complex and inefficient setups for generating these entangled photons, and relatively low efficiency of the entire system. In quantum metrology systems, maintaining a low photon number is crucial \cite{Slussarenko2017, Wolfgramm2013-ws, He2023, TAYLOR20161}. However, low brightness sources deeply degrade quantum imaging and sensing applications, due to long acquisition times of the measuring devices \cite{He2023}. Conversely, it is equally essential to achieve high resolution in real-time, which requires high-flux photon sources. Hence, the goal is to achieve the maximal photon flux possible while considering sample damage limitations. Furthermore, complex setups usually have unavoidable losses that prevent quantum metrology from fulfilling their full potential in addition to their large footprint \cite{Cimini:19}.

Recently, structured light \cite{Forbes2021} has been increasingly used to achieve better sensitivity in classical and quantum sensing schemes owing to its high dimensional nature, among them is optical rotation sensing. Optical rotation metrology is used for inertial navigation systems or for monitoring rotation in cryogenic environments \cite{doi:10.1126/sciadv.aau1540} among other applications. Structured light envisioned \cite{6990490, PhysRevA.83.053829} and enabled rotation sensing using twisted N00N states \cite{FicklerN00N}, polarization photonic gears \cite{D'Ambrosio2013}, and squeezed light \cite{10.1063/1.5066028,https://doi.org/10.1002/qute.202200055}. However, these studies faced challenges due to either inherent lossy systems that impeded the full potential of quantum light, lacked the measurement of mechanical rotation, relied on single photon sources (hence lacking entanglement enhancement), or were degraded while increasing the dimensions of structured light. Therefore, the combination of structured and quantum light still faces major challenges due to complex and lossy systems, hindering true quantum advantage.

In our study, we sense rotations utilizing a new concept we name \textbf{wavefront photonic gears} and a robust and high brightness source for directly generating N00N states with $N=2$ \cite{DiDomenico:22}. The method we present here addresses all the above issues of combining structured and quantum light: N00N state photons are generated in a compact source, and the setup that follows it uses very few elements with relatively low loss even for high-dimensional structured light. We demonstrate it by increasing the sensitivity  using higher dimensional structured light, while using short acquisition times, to measure rotation and angular acceleration, while retaining the coincidence count and visibility.

\begin{figure}[ht]
    \centering
    \includegraphics[width=1\linewidth]{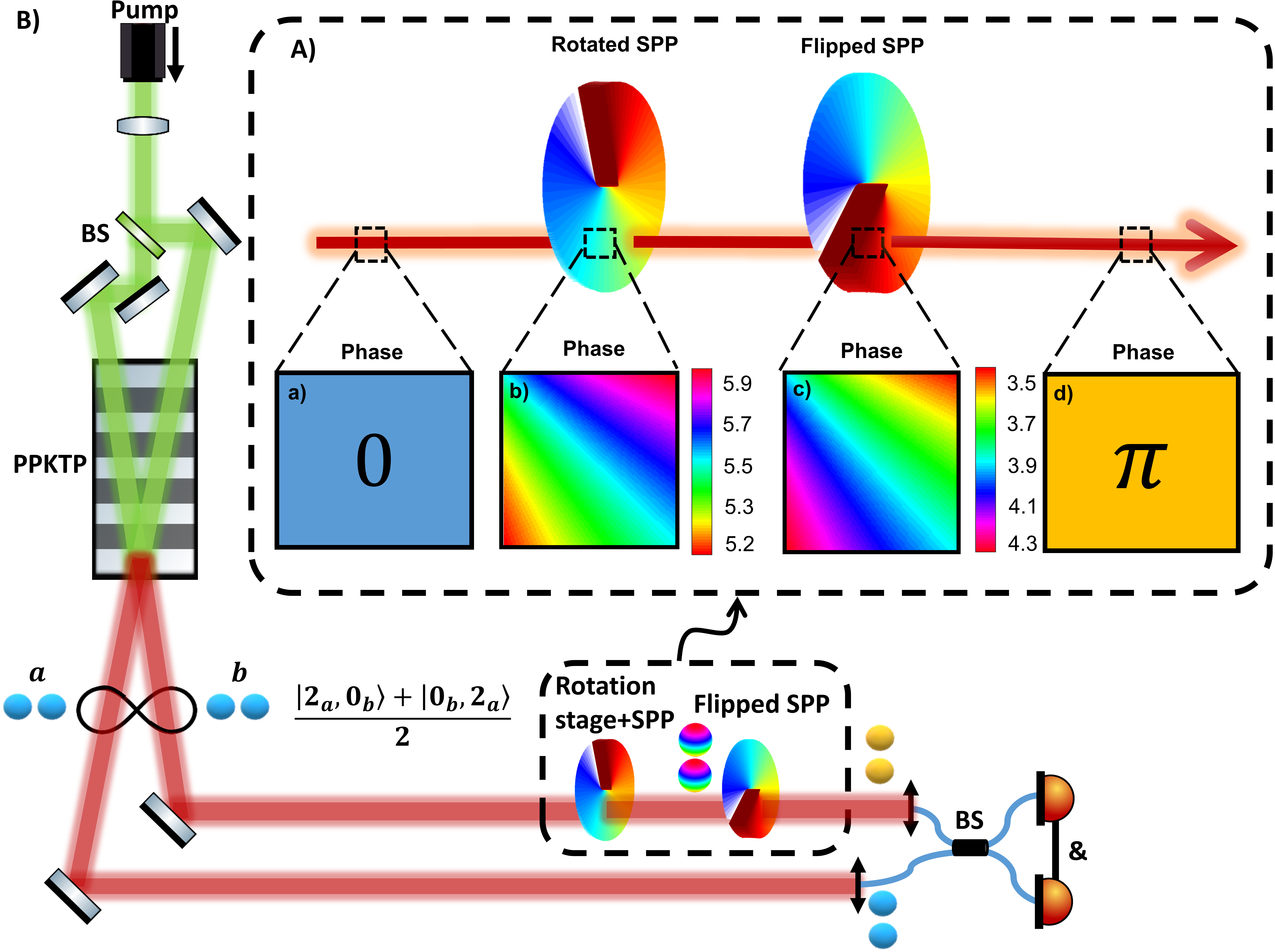}
    \caption{\textbf{Conceptual illustration of wavefront photonic gears and experimental set-up.} \textbf{A)} A collimated Gaussian mode with a constant phase (a) passes through a spiral phase plate with topological number $\ell=1$, accumulating a phase distribution (b). After a short propagation distance, the beam passes through a shifted reciprocal phase distribution (c), hence increasing the overall constant phase to $\pi$ in this example. (d).\textbf{B)} Experimental set-up. Pump shaping enables the generation of a two-photon path entangled N00N state immediately after the nonlinear crystal. Next, the setup includes an MZI with two oppositely oriented SPPs, one of which is mounted on a rotation stage. After interfering at the beam splitter (BS), coincidence detectors record data for measuring coincidence count. Blue particles represent photons with no wavefront phase shift, multi-colored particles represent structured photons, after the first SPP, and yellow particles represent an example of $\pi$ shifted wavefront photons for the $\ell=1$ case.}
    \label{fig:wavefront-photonic-gears}
\end{figure}

Using two spiral phase plates with opposite orientation, i.e., with topological charge of $\ell$ and $-\ell$, respectively, we convert mechanical rotations of a motor to wavefront phase shifts, hence allowing us to use interference to sense rotations, with $\ell$ fold improvement in the angular resolution. Spiral phase plates are typically aligned so that the beam passes through the center of the plate, e.g., to convert a Gaussian beam to a Laguerre-Gaussian beam \cite{Oemrawsingh:04}. However, the region near the singularity of the plate may distort the beam and scatter part of it, owing to fabrication limitations. For the rotation sensing application we present here, we overcome this limitation by sending the beam at some distance from the center of the spiral phase plate. By employing the N00N state source and coincidence detection in this system, we obtain an additional $N$-fold improvement in resolution. Hence, as will be shown below, the angular resolution scales as $N \times \ell$, and the sensitivity increases with $\ell$.

In addition to super-resolved rotation sensing, our high brightness source allows us to measure fast rotations (reaching 10 deg/sec), and angular acceleration (here measured at 1 deg/$\text{sec}^2$), which is a significant challenge in other, low brightness schemes. With further reductions of both losses, drift and improvements in detection efficiency, this scheme may enable us to approach unconditional violation of the shot noise limit for all the detected photons in the system \cite{PhysRevLett.98.223601, Slussarenko2017}.

\section{Theoretical model}

Upon impinging a spiral phase plate (SPP), the beam acquires a phase $\phi_1$ (Fig.1):
\begin{equation}
\phi_1(x,y) = \begin{cases}
    \arctan\left(\frac{y}{x}\right)\ell, &  x,y \in \Psi_0 \\
    0, & \text{otherwise}
\end{cases}
\end{equation} 
After impinging the second flipped SPP, the beam acquires the following phase $\phi_2$:
\begin{equation}
\phi_2(x,y) = \begin{cases}
    -\arctan\left(\frac{y}{x}\right)\ell + \theta\ell, & x,y \in \Psi_1 \\
    0, & \text{otherwise}
\end{cases}
\end{equation}
where $\Psi_1$ is a Gaussian beam with phase $\phi_1$ that is displaced from the second SPP singularity. $\theta$ is the rotation angle, and $\ell$ is the SPP topological charge. Hence, after two SPPs, the output beam $\Psi_2$ satisfies, classically: 
\begin{equation}
\text{phase}(\Psi_2(x)) = \text{phase}(\Psi_0(x)) + \theta\ell
\end{equation}

Therefore, the two SPPs serve as a wavefront photonic gear, converting mechanical rotation to flat phase delay, that is dependent on the topological charge of the SPP. We assume that the second SPP is in the near field of $\Psi_1$. Our scheme differs from other photonic gears implementations \cite{Barboza2022,D'Ambrosio2013} by eliminating the need for polarization handling. This feature enhances its versatility, enabling to improve the sensitivity in various interferometric configurations.

We now turn to analyze how wavefront photonic gears can be enhanced by quantum entanglement, here analyzed for the case of  N00N states with N=2 that will be denoted 2002. According to Eq.(3), a single arm of a Mach Zehnder interferometer (MZI) accumulates a phase of $\theta \ell$. Hence, after rotation of a single SPP in a single arm of the path entangled 2002 state, the state accumulates 2$\theta \ell$ according to $|\psi\rangle=|2_a,0_b\rangle +e^{i2\theta \ell}|0_a,2_b\rangle$ where a,b denote the two arms of the interferometer. Following \cite{PhysRevA.83.053829,FicklerN00N}, by mixing the output of the two arms using a 50:50 beam splitter, 
the final state before the detectors is\cite{DiDomenico:22}:

\begin{equation}
\begin{aligned} {|\psi \rangle _{2002}} &= \sin (\theta\ell) \frac{{|2 \rangle |0 \rangle + |0 \rangle |2 \rangle }}{{\sqrt 2}} + cos(\theta\ell) |1 \rangle |1 \rangle . \end{aligned}
\end{equation}

 Hence, by rotating one of the SPPs and measuring the coincidence (i.e. the $|1 \rangle |1 \rangle$ state), the signal oscillates $N\times\ell$ times for 360° rotation (in our specific case $2\times\ell$), thus exhibiting super-resolution.
This sets the following bound on the uncertainty (Heisenberg limit):

\begin{equation}
\Delta \theta =\frac{1}{\sqrt{M}N\ell}
\end{equation}

Where M is the number of measurements, which saturates the quantum Cramer-Rao bound \cite{TAYLOR20161}. Hence, it is of interest to use quantum entanglement and push the sensitivity further using the 1/N scaling, whereas the shot noise limited error scales as $1/\sqrt{N}$. Importantly, in our work, we benefit from all three multiplicative in the denominator in Eq.(5) in a manner that pushes the sensitivity in real life scenarios. We use N=2 using a bright N00N state source, hence increasing M. In addition, we increase the topological charge number and avoid aperture limitations, resulting in a negligible coincidence pair loss (the coincidence rate dropped by $4.8\%$ percent, from 44.9k/sec with $\ell$=1 to 42.7k/sec with $\ell$=16, shown in Fig.2) showcasing the potential of wavefront photonic gears with a minimal trade-off between larger $\ell$ to coincidence counts.

\section{Experimental setup}

The experimental setup is shown in Fig.1. 
The quantum source is a bright N00N state source that consists of a continuous wave 532.25 nm laser with output power of 75mW after the residual pump filter, a beam splitter and a 2 cm PPKTP crystal poled for type-0 interaction \cite{DiDomenico:22} held inside an oven with a temperature of 36 degrees Celsius. It is followed by a Mach-Zhender interferometer that utilizes the wavefront photonic gears mechanism, consisting of two opposite SPPs where the first one is mounted on a mechanical rotation stage. The two SPPs are positioned such that the beam passes roughly 2mm from the singularity point at the center of the SPPs. The two SPPs are located ~5cm from one another. The two N00N state paths are then coupled to two 1064nm single mode polarization maintaining fiber and interfere inside a 50:50 fiber beam splitter. Then, photons from the two arms are detected using two (SNSPDs) superconducting nanowire single photon detectors (Single Quantum Eos) with 81\% and 86\% efficiency.

\section{Rotation measurement}

We measured the coincidence at different angles for $\ell=1$ and $\ell$=16, that were set by the motor, and were equally spaced.
We analyzed the data using a weighted least squares fit, where each point was weighted by the reciprocal of the measured variance, from a set of 20 repeated measurements. The resulting fitted curve is as follows\cite{FicklerN00N}:

\begin{equation}
\frac{A}{2} \left[ 1 - \cos \left( \frac{ \pi N \ell}{180^\circ} \theta - C \right)\right] + B
\end{equation}

 Where $A$ is the amplitude of the cosine curve, $B$ is the offset and  $C$ is the parameter specifying the initial rotation point. Thus, the expression $\left[ \frac{A}{A + 2B} \right]$ provides an estimation of the curve's visibility, as determined by the fit.
Mechanical rotation displacement causes increased drift in the detected signal. Therefore, each of the 20 measurements commenced from a distinct relative phase between the two arms. To tackle this issue, we employed a cross-correlation analysis to identify the relative phase ('C') of each measurement. Then, we aligned each measurement correctly with the desired starting point on the curve.

\begin{figure}
   \hspace{-0.6cm}
    \includegraphics[width=1.1\linewidth]{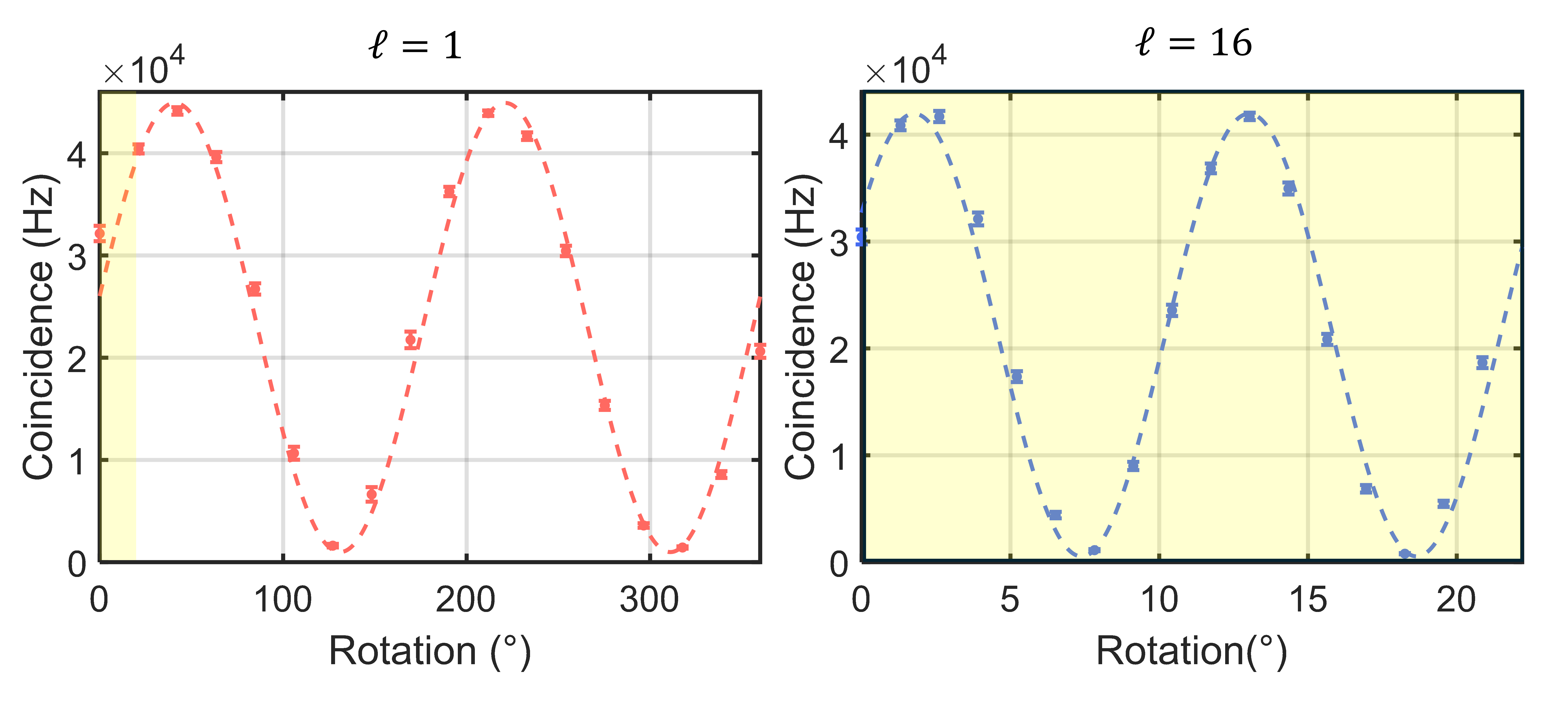}
    \caption {\textbf{Coincidence counts versus rotation angle plots for $\ell=1$ and $\ell=16$.} Both figures demonstrate that the detected signals exhibit super resolution. In the case of $\ell=16$, the coincidence count oscillates 16 times more over a 360-degree rotation than in the $\ell=1$ scenario (and both oscillate twice times more than the classical case). Each data point comprises 20 measurements taken from 20 different rotation experiments, all aligned using cross-correlation analysis. Shaded yellow area corresponds to the segment where the signal with $\ell=16$ performs two oscillations, for comparison.}
    \label{fig:enter-label}
\end{figure}

Fig. 2 shows the results of two experiments, one with $\ell=1$ and the other with $\ell=16$ with an acquisition time of 0.1 seconds and a coincidence window of 0.5ns, normalized to coincidence per second.  Clearly, the coincidence count exhibit super resolution due to quantum correlations, where in the $\ell=1$ the detected pairs oscillate  twice in a 360 degrees rotation, and in the $\ell=16$ the coincidence count oscillate 32 times (here we show almost two oscillations in 20.8° degrees). For each of these measurements, the visibility values are  $[95.7\%]$ and  $[97.6\%]$ respectively. 

We compute the angular uncertainty using the following equation \cite{FicklerN00N}:

\begin{equation}
\Delta \theta = \frac{\Delta M(\theta)}{ \frac{A N\ell\pi}{360^\circ}\left| \sin\left(\frac{\pi N \ell}{180^\circ} (\theta - C)\right) \right|}
\end{equation}

Where $\Delta M(\theta)$ represents the standard deviation for each measurement angle, calculated from 20 measurements. The angular precision results are presented in Fig.3 for $\ell = 1$ and $\ell = 16$. The lowest uncertainty for $\ell = 1$ ($\ell = 16$) is 2.28° (0.09°) degrees.

\begin{figure}
     \hspace{-0.6cm}
    \includegraphics[width=1.1\linewidth]{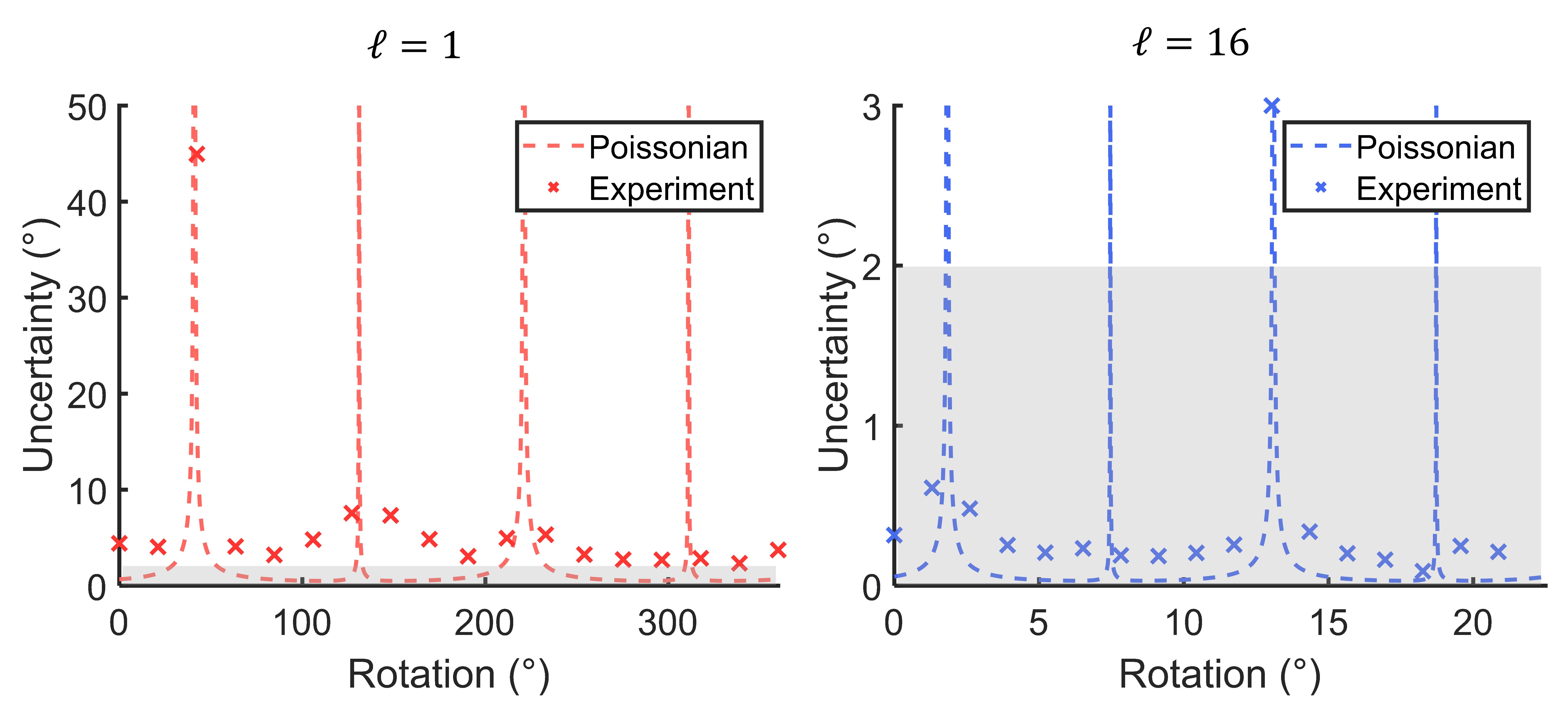}
    \caption{\textbf{Angular uncertainty for two cases: $\ell=1$ in red and $\ell=16$ in blue.} The dashed curves are calculated from Eq.(7) while taking into account Poissonian errors for $\Delta M(\theta)$, as determined from Eq.(6). The crosses represent experimentally determined precision values calculated using Eq.(7), where for this case $\Delta M(\theta)$ signifies the standard deviation of a single rotation measurement. Shaded area corresponds to the same uncertainty values, for comparison.}
    \label{fig:enter-label}
\end{figure}

Importantly, by utilizing larger values of $\ell$, we obtained a 25-fold experimental uncertainty reduction associated with $\ell=16$ compared to $\ell=1$, without changing the acquisition time and the number of coincidence counts. 

Increasing $\ell$ allows us to approach the Poissonian error regime (Fig.3) . Specifically, for $\ell=1$, the minimal uncertainty is 4.5 higher than the Poissonian errors regime. Conversely, for $\ell=16$, our measurements are 2.2 higher than the Poissonian errors regime, which is attributed to the decreased mechanical rotation for detecting a single oscillation. The gap between expected and desired results is primarily due to discrepancies arising from changes in coupling due to mechanical rotation, and interferometer drift during extended measurements using a physical motor.  This gap can be mitigated using a specially built mechanical rotor that will have less effect on the stability of the interferometer, and active interferometer stabilizers.

In the analysis above we considered only the coincidentally detected photons, but the single count rate for both detectors are roughly 500,000 and 490,000, compared to ~45,000 coincidence detections per second. To reach unconditional violation of the shot noise limit with two photon N00N states, the system efficiency and visibility must satisfy $\eta^{N}V^2N>1$,  where $\eta$ is the Klyshko efficiency (here measured $9\%$ ), and V is the visibility \cite{Slussarenko2017}. To reach this condition,   assuming near unity detection efficiency \cite{Reddy:20}, we need to increase the coupling efficiency, which constitutes the main loss mechanism, by a factor of ~3 (for example using objective lenses and alignment stages with tilt and rotation).

\section{Acceleration measurement}

\begin{figure}[!htb]
 
     \hspace{-1cm}
    \includegraphics[width=1.2\linewidth]{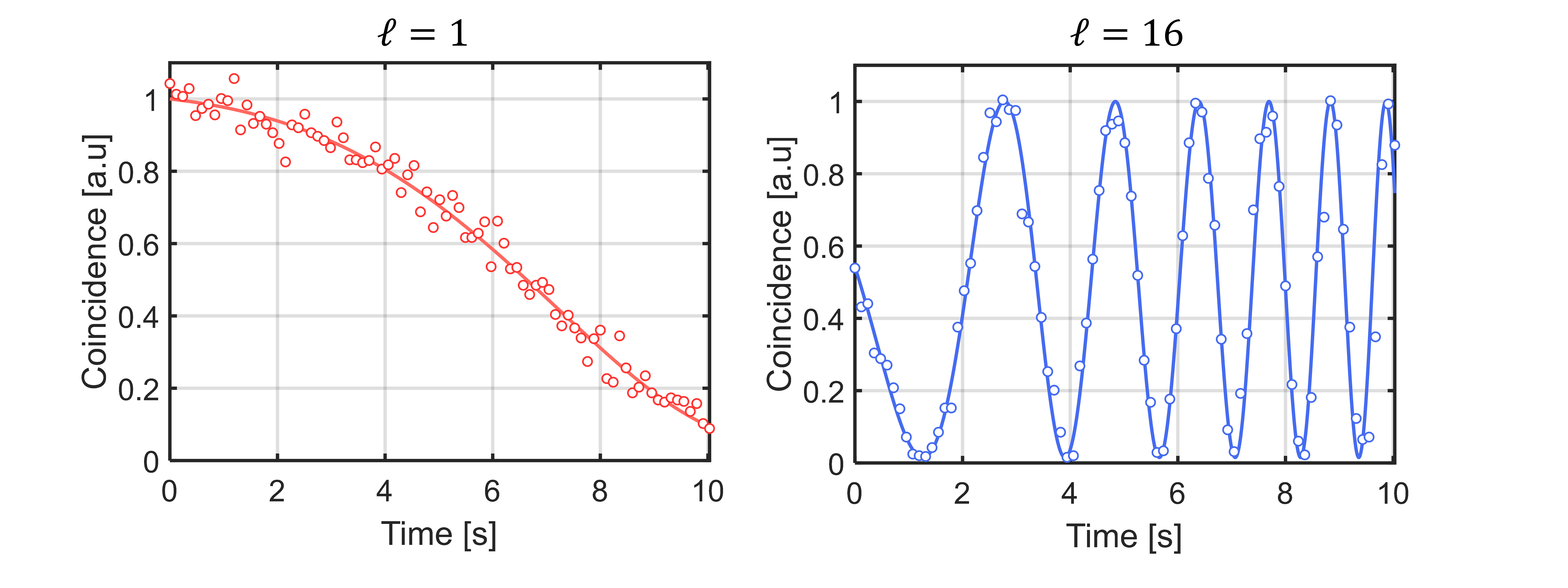}
    \caption{\textbf{Acceleration measurements.} Experimental (circles) and least square fitting function (continuous curves)  for a rotating motor under constant acceleration (1 deg/sec$^2$) over a 10-second duration. Two distinct topological charges, $\ell=1$ and $\ell=16$, were employed. Acceleration measurements were extracted from the fitted curves.}
    \label{fig:enter-label}
\end{figure}

 Notably, using our high-brightness scheme, we are not limited to slowly varying angular displacement but can measure fast rotations and angular acceleration, using structured  quantum light. Here we measured the acceleration of a rotating motor using an acquisition time of 0.01s. The motor was subjected to a constant acceleration of 1° deg/sec$^2$ and the measurement was limited to T=10 seconds.The measurements were executed using our wavefront photonic gear sensor, employing two distinct topological charges, $\ell$=1 and $\ell$=16. To estimate the given rotation, we fitted a chirp function to the detected coincidence pair count (Fig.4):

\begin{equation}
\frac{A}{2} \left[ 1 - \cos \left(\theta_0+ \frac{\pi N \ell}{180}w_0t + \frac k2 t^2\right)\right] + B
\end{equation}
where $k= \frac{\pi N \ell\left(w_f - w_0\right)}{180 T}$ is the acceleration parameter, from which the angular acceleration can be extracted,$\frac{\pi N \ell}{180}w_{0(f)}$ is the initial (final) angular frequency, and  $\theta_0$ is the initial phase. Using least squares fit for $\ell$=1, we registered an acceleration of 1.51° deg/sec$^2$ squared, which is quite far from the nominal angular acceleration of 1° deg/sec$^2$ squared, but for $\ell$=16, we obtained a measurement of $1.015^\circ \pm 0.021^\circ$ deg/sec$^2$. For $\ell$=1, the 10 second measurement duration proved inadequate for achieving a precise fit. This limitation arises due to the numerous potential fitting functions applicable in such circumstances. Therefore, the enhanced super-resolution provided by higher $\ell$ values and photon numbers N contribute to shorter measurement time and improved accuracy.

\section{Conclusions}

We demonstrated a new rotation sensing mechanism that uses quantum light and just two spiral phase plates. The small footprint of wavefront photonic gears makes it highly suitable for real-life scenarios. This scheme does not need any additional beam manipulation with respect to other MZI: Both the nonlinear crystal and the two phase plates are on -axis and merely put inside an already established MZI scheme (with an additional filter for the 532nm laser). Remarkably, no decrease in visibility and only a negligible reduction in the detected pair rate  is measured when using different topological charges of the SPPs. Hence we envision that the concept of wavefront photonic gears can be used with much higher topological charge SPPs, or even spiral phase mirrors \cite{PNAS10000} without major complications such as degrading beam quality. The reason is that usually, SPPs are manufactured with an unavoidable singularity point that is usually neglected for large beams. However, our system successfully eliminates this constraint, operating in close proximity to the singularity. The high brightness of our setup allowed us to use it for measuring angular acceleration and fast angular rotations that may help measure the rotation angle at any given angle using real-time feedback control \cite{Berni2015}. Furthermore, it may allow multi-photon sensing schemes, increasing the resolution using higher order photon pairs events. The concept of wavefront photonic gears is not limited to quantum light and can be also used with classical light, by simply using two phase plates and conventional interferometers. Additionally, the sensitivity may be increased by repeatedly passing through various locations on the same SPP, mimicking the advantages of multi-pass sensing schemes \cite{Israel:23}. Lastly, one can generalize our scheme to sense linear displacement in the same manner \cite{Barboza2022}, or to combine squeezed light with wavefront photonic gears \cite{PhysRevLett.130.070801}.

\begin{backmatter}
\bmsection{Funding}
Tel Aviv University Center for Quantum Science and Technology.; Israel Science Foundation ((969/22)); PAZY Foundation.

\bmsection{Disclosures}
The authors declare no conflict of interest.

\bmsection{Data Availability Statement}
All data, code, and materials used in the analysis are available upon reasonable request.

\end{backmatter}

\bibliography{sample}






\end{document}